\documentclass[12pt,reprint,eqsecnum,floats,aps,amsmath,amssymb,nofootinbib,prd,onecolumn,showpacs,superscriptaddress]{revtex4-1}

\usepackage{graphicx}
\usepackage{amsmath,amssymb}
\usepackage{hyperref}
\usepackage{graphicx}
\usepackage{subfigure}
\usepackage{arydshln}

\begin{document}

\title{Hamiltonian diagonalization in hybrid quantum cosmology}

\author{Beatriz Elizaga Navascu\'es}
\email{beatriz.b.elizaga@gravity.fau.de}
\affiliation{Institute for Quantum Gravity, Friedrich-Alexander University Erlangen-N{\"u}rnberg, Staudstra{\ss}e 7, 91058 Erlangen, Germany}

\author{Guillermo  A. Mena Marug\'an}
\email{mena@iem.cfmac.csic.es}
\affiliation{Instituto de Estructura de la Materia, IEM-CSIC, Serrano 121, 28006 Madrid, Spain}

\author{Thomas Thiemann}
\email{thomas.thiemann@gravity.fau.de}
\affiliation{Institute for Quantum Gravity, Friedrich-Alexander University Erlangen-N{\"u}rnberg, Staudstra{\ss}e 7, 91058 Erlangen, Germany}

\begin{abstract}

We explore the possibility of selecting a natural vacuum state for scalar and tensor gauge-invariant cosmological perturbations in the context of hybrid quantum cosmology, by identifying those variables for the description of the perturbations that display a dynamical behavior adapted in a specific way to the evolution of the entire cosmology. We make use of a canonical formulation of the whole of the cosmological system (background geometry and perturbations) in which the perturbative gauge-invariant degrees of freedom are identified as canonical variables. Introducing background-dependent linear canonical transformations that respect the spatial symmetries of the background on these perturbations and completing those canonical transformations for the entire system, we are able to characterize a generic collection of annihilation and creationlike variables that obey the dynamics dictated by a respective collection of Hamiltonians. We then impose that such Hamiltonians possess no self-interaction terms so that, in a Fock representation with normal ordering, they act diagonally on the basis of $n$-particle states. This leads to a semilinear first-order partial differential equation with respect to the background for the coefficients that define the annihilation and creationlike variables for all Fourier modes, as well as to a very precise ultraviolet characterization of them. Such first-order equation contains, in the imaginary part of its complex solutions, the complicated second-order field equation that typically arises for the time-dependent frequency of the perturbations in the context of quantum field theory in curved spacetimes. We check that the asymptotic knowledge acquired allows one to select the standard vacua in Minkowski and de Sitter spacetimes. Finally, we discuss the relation of our vacuum and the standard adiabatic vacua, and check that our asymptotic characterization of variables with a diagonal Hamiltonian displays the properties that would be desirable for an adiabatic state of infinite order.

\end{abstract}


\maketitle

\tableofcontents
\clearpage\thispagestyle{empty}

\section{Introduction}

Quantum field theory in curved spacetimes \cite{wald} has proven to be a powerful tool to extract predictions out of the presumable quantum fluctuations of fields that encode {\emph{small}} deviations around a known spacetime (often a highly symmetric one). A very remarkable example is the predicted statistical distribution of temperature anisotropies of the cosmic microwave background (CMB) \cite{planck,planck-inf,planck2}, starting from a Gaussian state for the inhomogeneities of our primordial Universe during an inflationary era, treated as perturbations \cite{mukh,langlois}. This Gaussian state can be regarded as the vacuum of a specific Fock representation of those perturbations, which in the context of inflation is typically identified as one of the most symmetric states in de Sitter spacetime, namely the Bunch-Davies vacuum \cite{BD}. This choice leads to a quantization of the perturbations with a particularly well-tamed dynamics, as the background cosmology on which the inhomogeneities propagate evolves in a quasi de Sitter way, according to the paradigm of slow-roll inflation within general relativity.

Despite the success of the use of a vacuum inspired by the Bunch-Davies proposal in predicting a power spectrum for the cosmological perturbations (constructed out of the two-point function) that leads to a distribution of temperature anisotropies matching the current observations of the CMB, one may still pose a few theoretical concerns about the robustness of this selection. On the one hand, this standard choice in cosmology is only adapted to a very specific type of dynamics of the scale factor of the homogeneous and isotropic background geometry, namely a period of quasi de Sitter accelerated expansion. One can wonder whether there might have been a previous non-inflationary epoch in the primordial stage of the Universe where the quantum evolution of the inhomogeneities, and even of the background homogeneous geometry itself, might have left traces that can be observed nowadays. In fact, it is reasonable to think that this may have been the situation if the quantum fluctuations of the non-perturbative homogeneous cosmological background can lead to an effective scenario where the classical relativistic Big Bang singularity is no longer present, so that one can safely extend the past evolution of the Universe beyond the onset of inflation \cite{AAN,Universe}. In any situation of this type, and more generally for an arbitrary evolution of the homogeneous scale factor, the choice of a vacuum state for the perturbations that is optimally adapted to the dynamics of the system remains an open issue. In the context of quantum field theory in curved spacetimes, a commonly accepted strategy to bypass this issue for scalar fields (which mathematically serve to describe the most relevant cosmological perturbations) is to restrict all considerations to the so-called adiabatic states \cite{adiabatic,adiabaticLR}, mainly owing to their ultraviolet, or short-scale, properties that allow for conventional regularization schemes of the stress-energy tensor associated with the inhomogeneous field under study \cite{adiabaticreg1,adiabaticreg2}. This connects with the second caveat that we would like to comment on here: the need to regularize or renormalize quantities that are constructed out of quadratic (and more generally of non-linear) functions of the field operators, such as the Hamiltonian that drives the dynamics of the field. This is a well-known problem of quantum field theory in curved spacetimes, that in particular is relevant in the simplest case of a field propagating on a homogeneous and isotropic background.

With the above two issues in mind, in this work we would like to discuss and propose physical arguments for the construction of a specific vacuum state of the cosmological perturbations and, in a more general context, for any scalar field minimally coupled to a spatially flat, homogeneous, and isotropic spacetime, requiring that this state displays well-behaved dynamics, as it is the case, for instance, of the Bunch-Davies vacuum in de Sitter spacetime. In order to do so, we will work within a canonical formulation for the complete system that is composed of the homogeneous and isotropic background cosmology and the inhomogeneous perturbations. The conceptual advantages of this canonical treatment are two-fold. On the one hand, we would like that our characterization of a quantization for the cosmological perturbations could be of use beyond the context of quantum field theory in curved spacetimes, and in particular in the framework of what has been called hybrid quantum cosmology \cite{hybrid1,hybrid2}. This is an approach for the quantization of inhomogeneous cosmological spacetimes that is based on a separation of the phase space of the system in two different sectors, one that encodes the homogeneous degrees of freedom and another one that captures the information about the inhomogeneities, for each of which one adopts different quantization techniques and representations. More specifically, under the assumption that the most relevant effects arising from a genuine theory of quantum gravity would likely affect the homogeneous sector in a more fundamental way, in the hybrid scheme one describes this sector with a representation inspired in quantum gravity. The remaining part of the phase space that contains the inhomogeneities is then quantized by means of some choice of Fock representation (to be fixed depending on the model). 

This hybrid approach was first introduced in connection with loop quantum cosmology \cite{lqc1,lqc2,ap}, a quantization scheme for the homogeneous degrees of freedom which is inspired by the non-perturbative and canonical program for the quantization of general relativity known as loop quantum gravity \cite{lqg}. However, in principle this hybrid philosophy is applicable as well to other canonical quantum cosmology techniques, as long as the resulting representation of the constrained Hamiltonian of the system can be  defined consistently as an operator in some Hilbert space. The separation of the degrees of freedom in phase space as homogeneous or inhomogeneous ones leads us to the second advantage of dealing with the entire cosmological system in a canonical way, in our objective of selecting a vacuum for the cosmological perturbations. Indeed, under the adopted hybrid perspective, it is of utmost importance to declare, in some physically meaningful way, which part of the system encodes the homogeneous information and which part describes the inhomogeneities, as these two types of degrees of freedom may in principle be mixed by means of canonical transformations in the entire cosmological system. In other words, from the perspective of the total Hamiltonian, there is freedom in splitting the total dynamics of the system and assigning the resulting parts to the two different sectors, using canonical transformations that mix them. A direct consequence is that one has freedom in defining an infinite collection of sets of annihilation and creationlike variables for the description of the inhomogeneities, such that their dynamics may vary from one set to another. Selecting one of those sets as a privileged choice by imposing certain physical criteria, e.g. a good quantum behavior of the Hamiltonian that generates their dynamics, then allows us to simultaneously fix the Fock vacuum for the inhomogeneities and, moreover, obtain a specific splitting of the phase space of the cosmological system that facilitates its hybrid quantization.

The hybrid approach has been extensively developed over the last years in the quantum treatment of homogeneous and isotropic cosmologies with a scalar inflaton field and perturbations of both the metric and the matter content \cite{hybr-inf1,hybr-inf2,hybr-inf3}. In particular, it has achieved a canonical formulation of the entire system (truncated at lowest nontrivial order in perturbations) purely in terms of perturbative gauge-invariant fields \cite{hybr-ref,hybr-ten}. These fields, even though they are not Dirac observables of the entire system, commute under Poisson brackets with the linear perturbations of the Hamiltonian and vector constraints (hence the given adjective for them), and are given by the scalar Mukhanov-Sasaki (MS) gauge invariant (for its exact definition, see Refs. \cite{sasa,kodasasa,mukhanov}) and the metric perturbations of tensor nature. In this work we will first analyze some properties of the Hamiltonian that generates the dynamics of these traditional gauge-invariant perturbations, when it is quantized following the hybrid prescription. We will see that it cannot be densely defined on any set of particle states arising from the possible choices of a Fock vacuum for the MS or tensor fields, namely from the possible choices of annihilation and creationlike variables obtained via linear background-independent canonical transformations of these fields. Therefore, if one wants to keep any hope to find a(n alternative) Hamiltonian for the perturbations with well-behaved quantum properties, further elaboration of the gauge invariants selected to describe the physical inhomogeneous degrees of freedom is in order, as well as a reconsideration of their dynamics, along the canonical line of attack that we have mentioned above. With this motivation, we will then characterize those sets of (gauge-invariant) annihilation and creationlike variables for the MS and tensor perturbations that evolve according to an alternate Hamiltonian which can be defined with the desired level of rigor as a Fock operator. Given the freedom that will still remain available in our constructions, we will then discuss the feasibility of a physically appealing choice, namely the selection of annihilation and creationlike variables for the perturbations with a Hamiltonian that has no self-interaction terms, terms which would be responsible of the dynamical production of pairs of excitations from the vacuum. After clarifying the relation between this type of choice and the standard representations allowed for the MS and tensor fields in the context of quantum field theory in curved spacetimes, we will finally discuss whether our proposed vacuum coincides with the Bunch-Davies vacuum, or with the Minkowskian one, when the background cosmology is restricted to de Sitter or to resemble flat spacetime, respectively. Moreover, in a more general context, we will compare our choice with the standard adiabatic states.

The detailed structure of this work is the following. In Sec. \ref{sec2} we first present the Hamiltonian formulation of the cosmological model on which we focus our analysis, namely a spatially homogeneous and isotropic flat spacetime minimally coupled to a scalar field with potential, both of which are perturbed with small inhomogeneities. Considering then the corresponding Hamiltonians for the gauge-invariant perturbations, derived from a quadratic perturbative truncation, we show that there exists no Fock representation of them with a well-defined action of the vacuum. Motivated by this, in Sec. \ref{sec3} we introduce new annihilation and creationlike variables for the description of the gauge invariants, obtained by means of background-dependent canonical transformations, and find the necessary and sufficient conditions on these variables so that their quantum Hamiltonians act properly on the vacuum state. Moreover, going further on the characterization of possible choices of variables, we determine the cases in which the Hamiltonians of the gauge-invariant perturbations are proportional to the number operator for each Fourier mode, paying special attention to the restrictions that this criterion imposes on the ultraviolet Fourier sector. With the resulting family of Fock representations at hand we discuss, at the end of that section, additional physical criteria to constrain the remaining freedom in the available choices. Next,  in Sec. \ref{sec4} we check that, when the evolution of the perturbations resembles that of a field in Minkowski or de Sitter spacetime, our asymptotic characterization serves to select the vacuum that indeed coincides with the highly symmetric choices that are standard in those situations. In Sec. \ref{sec5} we discuss the relation of our vacuum with other conventional choices for more general cosmological backgrounds. We first introduce the so-called adiabatic states and point out the typical obstructions that arise in their construction. We then compare them with our choice, inspired by Hamiltonian diagonalization, and argue the possible advantages of our approach. Finally, in Sec. \ref{concl} we summarize and comment our results. Throughout the work, we set the reduced Planck constant and the speed of light equal to one, $\hbar=c=1$.

\section{Mukhanov-Sasaki and tensor Hamiltonians}\label{sec2}

We start by considering a standard Friedmann-Lema\^itre-Robertson-Walker (FLRW) cosmological model that allows for the presence of small (perturbative) inhomogeneities and an inflationary epoch within Einstein's theory. Because the non-perturbative part of this metric is highly symmetric under spatial transformations, it is most convenient to expand the inhomogeneous perturbations in eigenmodes of the Laplace-Beltrami operator associated with the homogeneous spatial part of the metric. We further restrict this homogeneous {\emph{background}} geometry to be spatially flat and topologically compact. It can then be foliated in spatial hypersurfaces that are isomorphic to three-tori, with a compactification length that we call $l_0$. The corresponding Laplace-Beltrami operator is just the Euclidean Laplacian and, thanks to the compactness of the spatial hypersurfaces, its spectrum is discrete, formed by wave numbers $k$ equal to the Euclidean norm of wave vectors $\vec{k}$ such that $l_0\vec{k}/( 2\pi ) \in \mathbb{Z}^3$. One can then expand the metric and the scalar field in the eigenmodes of this Laplacian and isolate the purely inhomogeneous degrees of freedom, which are treated as small perturbations in this context, from the homogeneous ones, which, on their own, would describe a FLRW cosmology with a homogeneous scalar field as matter content. Finally, according to their transformation properties under the isometries of the spatial sections of the homogeneous background, the perturbations (and thus their mode expansions) can be characterized as scalar, tensor, or vector perturbations \cite{mukh}.

We regard the inhomogeneities as perturbations by truncating the Einstein-Hilbert action of the system at the lowest non-trivial polynomial order in all of their mode coefficients, namely at quadratic order. At this order of truncation, the vector perturbations are seen to be pure gauge and can thus be ignored in the treatment. In a canonical framework, it is possible to find a description of the truncated system in terms of only perturbative gauge invariants \cite{hybr-ref,hybr-ten}, with a total Hamiltonian that is the following linear combination of constraints:
\begin{align}
N_{0}\left[H_{|0}+{^sH_{|2}}+{^T{H}_{|2}}\right]+\sum_{\vec{k}\neq 0} l_{\vec{k}}H_{|1}^{\vec{k}}+\sum_{\vec{k}\neq 0} m_{\vec{k}}H_{\_ 1}^{\vec{k}}
\end{align}
where $N_{|0}$ describes the zero-mode of the lapse function, $l_{\vec{k}}$ describe the mode coefficients of its inhomogeneous perturbations, and $m_{\vec{k}}$ the mode coefficients of the shift vector \cite{hybr-ref}. Besides, $H_{|0}$ is formally identical to the constraint that one would find for the FLRW unperturbed cosmology:
\begin{align}\label{H0}
H_{|0} = \frac{1}{2l_0^3 a^3}\left[\pi_\phi^2-\frac{4\pi G}{3}a^2\pi_a^2+2l_0^6 a^6 V(\phi)\right], 
\end{align}
where $(a,\pi_a)$ and $(\phi,\pi_\phi)$ denote respectively the scale factor and the inflaton field, together with their canonical momenta, for the homogeneous cosmological background (and after being corrected with the addition of certain terms quadratic in perturbations, so that they form a canonical set with the perturbative gauge invariants \cite{hybr-ref}). In addition, $V(\phi)$ is the inflaton potential. On the other hand, ${^sH_{|2}}$ and ${^T{H}}_{|2}$ are, respectively, the MS and tensor Hamiltonians, that are specifically given by:
\begin{align}
\label{MSham}
&{^sH_{|2}}=\frac{1}{2a}\sum_{{\vec k} \neq 0} \left[\big( k^2 + s^{(s)}\big) |v_{\vec{k}}|^2 + |\pi_{v_{\vec{k}}}|^2\right], \\  \label{tensham}
&{^T{H}_{|2}}= \frac{1}{2a}\sum_{{\vec k} \neq 0,\epsilon} \left[\big(k^2+s^{(t)} \big)|d_{\vec {k},\epsilon}|^2+|\pi_{ d_{\vec {k},\epsilon}}|^2\right],
\end{align}
where the canonical pairs $(v_{\vec{k}},\pi_{v_{\vec{k}}})$ and $(d_{\vec {k},\epsilon},\pi_{ d_{\vec {k},\epsilon}})$ refer to the mode coefficients of the MS and tensor gauge-invariant fields, respectively. The reality of these fields is made manifest with the requirements $v_{-\vec{k}}=\bar{v}_{\vec{k}}$ and $d_{-\vec{k}}=\bar{d}_{\vec{k}}$, where the overbar denotes complex conjugation, as well as with analogous relations for their canonical momenta. Besides, $\epsilon$ is a dichotomic label that accounts for the two possible polarizations of the tensor perturbations, and $s^{(s)}$ and $s^{(t)}$ play the role of background-dependent masses, which are given by
\begin{align} \label{tmasses}
s^{(t)}=\frac{16\pi^2 G^2}{9l_0^6 a^2} \pi_{a}^2-8\pi Ga^2 V(\phi), \qquad
s^{(s)}= s^{(t)}+\mathcal{U},
\end{align}
with the MS potential $\mathcal{U}$ being equal to
\begin{align}
\mathcal{U}=a^{2}\left[V_{,\phi\phi}(\phi)+48\pi GV(\phi)-12 \frac{\pi_{\phi}}{a\pi_{a}}V_{,\phi}(\phi)-\frac{72l_0^6}{\pi_{a}^2} a^{4}V^{2}(\phi)  \right].
\end{align}
The sum of $H_{|0}$, ${^sH_{|2}}$, and ${^T{H}_{|2}}$ forms the zero-mode of the Hamiltonian constraint. Finally, $H_{|1}^{\vec{k}}$ and $H_{\_ 1}^{\vec{k}}$ are, respectively, Abelianized versions of the linearized Hamiltonian and diffeomorphism constraints, that generate the perturbative gauge transformations of the system at the considered order of truncation.

\subsection{Problems with the MS and tensor Hamiltonians in the hybrid approach}

The hybrid approach to the quantum description of the cosmological perturbations would rest on a suitable choice of a representation for the canonical degrees of freedom of the homogeneous sector inspired by quantum gravity, and of a Fock representation for the perturbative gauge invariants, together with a procedure to combine them. In this framework, the Abelianized linear perturbative constraints can be easily handled in the quantum theory: we impose them by simply requiring that the physical wave functions do not depend on the canonical momenta of these constraints, since those momenta are just pure gauge. Therefore, the first non-trivial problem that one has to face in order to obtain physical states of the quantum cosmology is the imposition of the operator that represents the zero-mode of the Hamiltonian constraint. In what concerns its properties as an operator, a reasonable assumption for any decent representation of the homogeneous cosmology inspired in a quantum gravity formalism is that one can define a well-behaved operator for $H_{|0}$, as well as for the homogeneous background observables that couple to the perturbations in the MS and tensor Hamiltonians. Under this hypothesis, one can argue that a correct definition of the Hamiltonian constraint demands that, when the homogeneous degrees of freedom are regarded as fixed quantities, as it is the case in the context of quantum field theory in curved (classical) spacetimes, the operators that represent the MS and tensor Hamiltonians can be properly defined on the vacuum associated with the Fock representation adopted for the perturbations. In particular, this guarantees that these Hamiltonians have a well-defined action on the dense Fock subspace spanned by $n$-particle states.

As it is typical in the Fock quantization of any field theory, there is an infinite ambiguity in the choice of representations for the MS and tensor gauge invariants. This freedom reflects the liberty that one has to select the variables that are directly promoted to quantum annihilation and creation operators acting on the cyclic vacuum state \cite{wald}. These variables, that we call annihilation and creationlike, can be defined in flat FLRW cosmology by linear combinations of the Laplace mode coefficients of the considered field and its momentum. Specifically, given any such annihilationlike variable, the corresponding creationlike one is set to be its complex conjugate. The linear combinations that define this annihilationlike variable must then be such that it has a nonvanishing Poisson bracket only with its creationlike partner, this Poisson bracket being equal to $-i$.
A natural way to restrict the representations of physical interest is to use the symmetry transformations of the system, imposing the invariance of the vacuum under them. In the case that we are studying, we see that the MS and tensor Hamiltonians only couple mode coefficients associated with the same tuple $\vec{k}$, and with the same polarization in the case of tensor modes. Besides, the functions that multiply the perturbative degrees of freedom in the Hamiltonian only depend on the wave vector $\vec{k}$ through its wave number $k$. In this sense, this contribution is independent of the degeneracy of each Laplace-Beltrami eigenvalue. For the tensor perturbations, furthermore, the coefficients of the perturbation variables in the Hamiltonian are insensitive to changes in the polarization. These symmetries in the relevant constraint of the system are, at least for the most part, a manifestation of the isometries of the homogeneous cosmological background. One can then realize that the annihilation and creationlike variables that lead to a vacuum that is invariant under these symmetries have the form
\begin{align}\label{constanni}
a_{\vec{k}}=f_k v_{\vec{k}} + g_k {\bar{\pi}}_{v_{\vec{k}}}, \qquad {\bar{a}}_{\vec{k}}={\bar{f}}_{k} {\bar{v}}_{\vec{k}}+{\bar{g}}_{k}\pi_{v_{\vec{k}}}
\end{align}
for the MS field, adopting analogous expressions for the tensor field. In principle, $f_k$ and $g_k$ are constant (i.e., time independent) coefficients that must satisfy
\begin{align}\label{sympl}
f_k {\bar{g}}_{k}-g_k {\bar{f}}_{k}=-i ,
\end{align}
so that the linear transformation that defines the annihilation and creationlike variables is canonical and leads to the correct Poisson algebra, namely $\{a_{\vec{k}},{\bar{a}}_{\vec{k}'}\}=-i\delta_{\vec{k},\vec{k}'}$. In terms of these variables, the MS Hamiltonian reads
\begin{align}\label{MShamaa}
{^sH_{|2}}=\frac{1}{2a}\sum_{{\vec k} \neq 0}\left\lbrace 2\bigg[\big(k^2 + s^{(s)}\big)|g_k|^2 + |f_k|^2\bigg]{\bar{a}}_{\vec{k}} a_{\vec{k}} - \bigg[\big(k^2 + s^{(s)}\big){\bar{g}}_k^2 + {\bar{f}}_k^2\bigg]a_{\vec k}a_{-\vec k} -\bigg[\big(k^2 + s^{(s)}\big)g_k^2 + f_k^2\bigg]{\bar{a}}_{\vec k}{\bar{a}}_{-\vec k} \right\rbrace.
\end{align}
A completely analogous expression is obtained for the tensor Hamiltonian by just replacing the mass $s^{(s)}$ with $s^{(t)}$, and the annihilation and creationlike variables with their counterparts for the tensor perturbations. A Fock representation of this Hamiltonian for fixed background cosmology can be obtained by promoting to annihilation and creation operators the variables $a_{\vec k}$ and ${\bar{a}}_{\vec k}$, respectively, imposing normal ordering on the non-commuting terms ${\bar{a}}_{\vec{k}} a_{\vec{k}}$. The resulting operator ${^s{\hat{H}}^{(F)}_{|2}}$ manifestly displays the destruction and production of pairs of excitations for an infinite collection of modes, a phenomenon caused by the products $a_{\vec k}a_{-\vec k}$ and ${\bar{a}}_{\vec k}{\bar{a}}_{-\vec k}$. These self-interaction terms in the Hamiltonian can lead to an ill-defined action on the vacuum state determined by the chosen set of annihilation and creation operators. Indeed, one can readily check that the image of the vacuum under ${^s{\hat{H}}^{(F)}_{|2}}$ has a finite norm in Fock space if and only if
\begin{align}\label{MSwelldef}
\sum_{{\vec k} \neq 0} \left|\big(k^2 + s^{(s)}\big)g_k^2 + f_k^2 \right|^2<\infty.
\end{align}

The convergence of this sum depends on the asymptotic behavior of the associated non-squared sequence, in the limit $k\rightarrow\infty$. If we assume that the mass $s^{(s)}$ can be treated as a continuous function of the homogeneous background variables, this behavior depends in turn only on the asymptotic properties of the sequences of constants $f_k$ and $g_k$, which are restricted by condition \eqref{sympl}, so that the imaginary part of $f_k \bar{g}_k$ equals $-1/2$, and hence $|f_k g_k|\geq 1/2$. It is not difficult to realize that, if $|f_k|$ is of the same asymptotic order as $|g_k|$ or smaller, this restriction implies that $|g_k|$ is of the asymptotic order of the unit or greater, and the sum \eqref{MSwelldef} immediately diverges. So the only admissible possibility is that $|g_k|$ is asymptotically negligible compared to $|f_k|$. If this is the case, then condition \eqref{sympl} implies that $|f_k|$ must grow when $k$ tends to infinity. Therefore, the only way in which the above sum can converge is that the term $f_k^2$ gets cancelled by $k^2 g_k^2$ asymptotically, up to subdominant contributions. Such cancellation is only consistent with condition \eqref{sympl} if
\begin{align}\label{fgasymp}
kg_k=if_k + \xi_k, \qquad |f_k|^2=\frac{k}{2}+o(k), \qquad \xi_k=o(k^{1/2}),
\end{align}
where the symbol $o(.)$ stands for asymptotically negligible compared to its argument, as $k\rightarrow\infty$. The elements of the studied sequence then become
\begin{align}\label{seq1}
\big(k^2 + s^{(s)}\big)g_k^2 + f_k^2=\big(2if_k \xi_k +\xi_k^2\big)\left(1+\frac{s^{(s)}}{k^2}\right)-f_k^2 \frac{s^{(s)}}{k^2}
\end{align}
and thus, since $s^{(s)}$ is time-dependent (in general, via a dependence on the background) while $f_k$ and $g_k$ are constant by definition, at arbitrary time the studied elements are generically at least of the asymptotic order of $f^2_k s^{(s)}k^{-2}$ in norm, which is $k^{-1}$, a fact that implies the non-square summability of the sequence over all wave vectors $\vec{k}\neq 0$. We thus conclude that, both for a fixed background cosmology or in the context of a hybrid quantization with the hypotheses that we have mentioned, the MS Hamiltonian cannot be defined with normal ordering as an operator that has a well-defined action on the Fock vacuum, and therefore it can neither be defined in the dense subspace spanned by the $n$-particle states constructed from it.

A completely analogous result is obtained for the tensor Hamiltonian, for which the only difference with respect to the MS Hamiltonian (apart from the sum over polarizations) is the explicit expression of the time-dependent mass. In fact, the same type of Hamiltonian with again only a distinct time-dependent mass can be found in the more paradigmatic example of a test Klein-Gordon field $\varphi$ minimally coupled to a classical FLRW cosmology with scale factor $a$. In order to derive it, it suffices to consider the time-dependent canonical transformation of the canonical pair $(\varphi,\pi_\varphi)$ to $(\tilde\varphi,\pi_{\tilde\varphi})$, with $\tilde\varphi=a\varphi$ and $\pi_{\tilde\varphi}=a^{-1}\pi_\varphi+a^{\prime}\varphi$, where the prime denotes the derivative with respect to the conformal time $\eta$ \cite{unitarity1,unitarity2}. Thus, the results of this section and of the following ones apply also to this scenario of quantum field theory in curved spacetimes, with the homogeneous background properly fixed as a solution of the linearized Einstein's equations.

\section{New perturbative gauge invariants and vacua}\label{sec3}

The main reason why we have found problems to define a physically interesting Hamiltonian operator for the MS and tensor perturbations is that the coefficients $f_k$ and $g_k$ do not depend on the background degrees of freedom, while the masses $s^{(s)}$ and $s^{(t)}$ do. One may wonder whether the introduction of a dependence of $f_k$ and $g_k$ on the variables $(a,\pi_a)$ and $(\phi,\pi_\phi)$ may make possible that one can arrive at a well-defined Hamiltonian operator on the Fock vacuum for the cosmological perturbations within the hybrid approach. The resulting annihilation and creationlike variables obtained through the linear transformation \eqref{constanni}, after allowing a background dependence of the coefficients $f_k$ and $g_k$, would clearly still describe gauge-invariant perturbative degrees of freedom, but with a different dynamical behavior from that of the standard MS and tensor variables. Actually, since their definition involves explicitly the homogeneous degrees of freedom, these new annihilation and creationlike variables no longer form a canonical set with the variables that we were using to describe that homogenous sector. If one wants to reach again a complete canonical set, something that is definitely convenient from a hybrid point of view, it is necessary to introduce some specifically modified homogeneous variables \cite{hybr-ref}. In more detail, the procedure to obtain these new variables for the description of the homogeneous sector is to consider the symplectic potential of the entire truncated system, which reads:
\begin{align}
\pi_a \text{d}a+ \pi_{\phi} \text{d}\phi + \sum_{\vec{k}\neq 0} \pi_{v_{\vec{k}}}\text{d}v_{\vec{k}} + \sum_{\vec{k}\neq 0, \epsilon} \pi_{d_{\vec{k},\epsilon}}\text{d} d_{\vec{k},\epsilon}
\end{align}
and then find a set of modified pairs $(\tilde{a},\pi_{\tilde{a}})$ and $(\tilde{\phi},\pi_{\tilde{\phi}})$ such that, up to an exact differential and truncating at quadratic order in perturbations, the symplectic potential becomes:
\begin{align}
\pi_{\tilde{a}} \text{d}\tilde{a}+ \pi_{\tilde{\phi}} \text{d}\tilde{\phi} +i \sum_{\vec{k}\neq 0} \bar{a}^{(s)}_{\vec{k}}\text{d}{a}^{(s)}_{\vec{k}} + i\sum_{\vec{k}\neq 0, \epsilon} \bar{a}^{(t)}_{\vec{k},\epsilon}\text{d}{a}^{(t)}_{\vec{k},\epsilon},
\end{align}
where, for the sake of clarity, we have emphasized with the superscripts $(s)$ and $(t)$ the dependence on the scalar and tensor degrees of freedom, respectively. For the class of annihilation and creationlike variables considered here, this procedure uniquely leads to the introduction of the following modified homogeneous variables:
\begin{align}
\tilde{x}=x-\Delta x ^{(s)}-\Delta x ^{(t)}, \qquad \pi_{\tilde{x}}=\pi_x-\Delta \pi_x^{(s)}-\Delta \pi_x^{(t)},\qquad x=a,\phi,
\end{align}
where we have defined
\begin{align}
&\Delta x=\frac{1}{2}\sum_{{\vec k}\neq 0}\left[2\text{Re}(\bar{g}_k \partial_{\pi_x}f_k-\bar{f}_k \partial_{\pi_x}g_k){\bar{a}}_{\vec{k}} a_{\vec{k}} +(\bar{f}_k\partial_{\pi_x}\bar{g}_k -\bar{g}_k\partial_{\pi_x}\bar{f}_k)a_{\vec k}a_{-\vec k} +(f_k\partial_{\pi_x}g_k -g_k\partial_{\pi_x}f_k){\bar{a}}_{\vec k}{\bar{a}}_{-\vec k}\right], \\ \nonumber &
\Delta {\pi_x}=-\frac{1}{2}\sum_{{\vec k}\neq 0}\left[2\text{Re}(\bar{g}_k \partial_{x}f_k-\bar{f}_k \partial_{x}g_k){\bar{a}}_{\vec{k}} a_{\vec{k}} +(\bar{f}_k\partial_{x}\bar{g}_k -\bar{g}_k\partial_{x}\bar{f}_k)a_{\vec k}a_{-\vec k} +(f_k\partial_{x}g_k -g_k\partial_{x}f_k){\bar{a}}_{\vec k}{\bar{a}}_{-\vec k}\right].
\end{align}
Here, $\text{Re}$ denotes the real part and we have dropped the superscripts $(s)$ and $(t)$ in the notation for simplicity, a convention that we will mantain in the following except for the time-dependent scalar and tensor masses. A sum over the two polarizations would be present in the modifications to the homogeneous degrees of freedom arising from tensor perturbations. At quadratic order of perturbative truncation in the action, these new homogeneous background variables ${\tilde a}$, $\pi_{\tilde{a}}$, $\tilde{\phi}$, and $\pi_{\tilde\phi}$ then form a canonical set with the annihilation and creationlike variables \eqref{constanni} defined with background-dependent coefficients $f_k$ and $g_k$.

At the adopted quadratic perturbative order, it is straightforward to calculate the new zero-mode of the Hamiltonian constraint. Expressed in terms of the new tilded background variables, this constraint is
\begin{align}
\left[H_{|0}+(a-\tilde{a}) \,\partial_{a}H_{|0}+(\pi_a-\pi_{\tilde{a}})\, \partial_{\pi_a}H_{|0}+(\phi-\tilde{\phi}) \, \partial_{\phi}H_{|0}+(\pi_\phi-\pi_{\tilde{\phi}}) \,\partial_{\pi_\phi}H_{|0}+ {^sH_{|2}}+ {^TH_{|2}}\right]({\tilde a},\pi_{\tilde{a}},\tilde{\phi},\pi_{\tilde\phi}),
\end{align}
where the final parentheses indicate functional evaluation of the terms inside the square brackets directly in the new homogeneous variables, i.e., evaluating the original dependence on the FLRW cosmological sector as if one identified the old and the new homogeneous variables. In consonance with this notation, in what follows we will write all our relevant quantities in the Hamiltonian constraint as functions of the old background variables, keeping in mind that, in the end, they must be identified with the new canonical ones. We call ${^s{\tilde H}_{|2}}$ and ${^T{\tilde H}_{|2}}$, respectively, the contributions to this Hamiltonian constraint that are quadratic in the MS and tensor perturbations. After multiplication by the homogeneous lapse, they can be interpreted in our truncated system as the generators of the classical linearized dynamics of the tensor and MS annihilation and creationlike variables defined with the background-dependent coefficients $f_k$ and $g_k$. We will therefore refer to them as the new MS and tensor Hamiltonians. In fact, whenever one can regard the homogeneous background as a fixed external entity, it is easy to check that the difference between the new and old Hamiltonians for the perturbations are just the corrections that arise owing to the fact that the new annihilation and creationlike variables are obtained by means of an explicitly time-dependent canonical transformation \cite{langlois,hybr-ref}. The expression of the new Hamiltonian for the MS gauge invariant is then
\begin{align}\label{newMSH}
&{^s{\tilde H}_{|2}}={^sH_{|2}}+\frac{1}{2}\sum_{{\vec k} \neq 0}\bigg[2\text{Re}\left({\bar f}_k\{g_k,H_{|0}\}-{\bar g}_k\{f_k,H_{|0}\}\right){\bar{a}}_{\vec{k}} a_{\vec{k}} + \left({\bar{g}}_k\{{\bar{f}}_k,H_{|0}\}-{\bar{f}}_k\{{\bar{g}}_k,H_{|0}\}\right)a_{\vec k}a_{-\vec k} +\text{H.c.} \bigg],
\end{align}
and a completely similar formula defines the new tensor Hamiltonian ${^T{\tilde H}_{|2}}$. Here, H.c. stands for Hermitian (complex) conjugate. Recall that $\{.,.\}$ are the Poisson brackets associated with our truncated system. We can now study whether the Fock representations of these new Hamiltonians for the perturbations, ${^s{\hat{\tilde H}}^{(F)}_{|2}}$ and ${^T{\hat{\tilde H}}^{(F)}_{|2}}$, with normal ordering and for fixed background (or in the hybrid quantization approach), can be defined properly on the vacuum state. As before, this will happen for ${^s{\hat{\tilde H}}^{(F)}_{|2}}$ if and only if
\begin{align}\label{MSwelldefN}
\sum_{{\vec k} \neq 0}\left| \big(k^2 + s^{(s)}\big)g_k^2 + f_k^2+af_k\{g_k,H_{|0}\}-ag_k\{f_k,H_{|0}\} \right|^2<\infty.
\end{align}

In order to analyze the asymptotic behavior at large $k$ of the associated non-squared sequence in a controllable manner, we restrict our considerations to functions of the wave number $f_k$ and $g_k$ with asymptotic expansions such that their Poisson brackets with $H_{|0}$ respect the asymptotic order of each term of those expansions. Then, arguments similar to those in the previous section show that a necessary condition for the convergence of the sum \eqref{MSwelldefN} is again that relation \eqref{fgasymp} holds asymptotically. The elements of the studied sequence turn out to be equal to the right hand side of Eq. \eqref{seq1} plus the additional contribution
\begin{align}\label{Poissoncorr}
\frac{a}{k}\left(f_k\{\xi_k,H_{|0}\}-\xi_k\{f_k,H_{|0}\}\right),
\end{align}
which is negligible compared to $f_k\xi_k$ with our assumptions. Recall now that the term with the factor $f_k^2k^{-2}$, which is proportional to the MS mass, is not square summable by its own, and therefore, if one requires the summability of \eqref{MSwelldefN}, it must be compensated with the other possibly dominant term, which is proportional to $f_k \xi_k$ and that can be background-dependent now. It then follows that another necessary condition for a correct definition of the new MS Hamiltonian on the vacuum is that, asymptotically,
\begin{align}\label{xicond}
\xi_k=-\frac{i}{2k^2}f_ks^{(s)}+\Gamma_k,\qquad \Gamma_k=o(k^{-3/2}).
\end{align}

If this restriction is imposed, the only summand in the considered sequence that might be not square summable over all $\vec{k}\neq 0$ is a term proportional to $f_k\Gamma_k$, term that behaves asymptotically as $k^{1/2}\Gamma_k$ according to Eq. \eqref{fgasymp}. Therefore, the desired summability condition \eqref{MSwelldefN} is finally satisfied if and only if
\begin{align}\label{Gammacond}
\sum_{{\vec k} \neq 0}k|\Gamma_{k}|^2<\infty.
\end{align}
A completely analogous analysis can be carried out for the new tensor Hamiltonian, replacing the time-dependent mass $s^{(s)}$ with $s^{(t)}$.

We conclude from the results in this section  that, by allowing the introduction of gauge-invariant annihilation and creationlike variables for the scalar and tensor perturbations that display a different dynamics than the standard MS and tensor variables, it is possible to find a Fock representation with normal-ordered Hamiltonians that are well defined on the dense subspace spanned by the $n$-particle states. The necessary and sufficient conditions for this to happen is that the background-dependent functions $f_k$ and $g_k$ that define these annihilation and creationlike variables, and satisfy Eq. \eqref{sympl}, are asymptotically constrained by relations \eqref{fgasymp}, \eqref{xicond}, and \eqref{Gammacond}, in the limit $k\rightarrow\infty$.

\subsection{Diagonalization}

Despite our result that the Hamiltonian operators ${^s{\hat{\tilde H}}^{(F)}_{|2}}$ and ${^T{\hat{\tilde H}}^{(F)}_{|2}}$ for fixed background can be defined on the dense Fock subspace of $n$-particles states, it turns out that they do not leave this domain invariant. This is because the image of an $n$-particle state (including the vacuum) under any of these operators contains an infinite linear combination of particle states, arising from the self-interaction terms in the Hamiltonians that destroy and produce pairs of excitations. This peculiarity complicates enormously the task of imposing quantum mechanically the zero-mode of the Hamiltonian constraint in the hybrid approach. Furthermore, we see that the conditions required for a proper definition of the Hamiltonian operators for the perturbations still allow for a huge ambiguity in the choice of annihilation and creationlike variables, inasmuch as  the function $\Gamma_k$ is only restricted by Eq. \eqref{Gammacond}. One can then ask whether it is possible to fix the remaining freedom in this choice of variables in such a way that the resulting Hamiltonians for the perturbations become proportional to the number operator. Most likely, this would alleviate both of the problems listed above, as it would eliminate all the self-interaction terms in the Hamiltonians, and much (if not practically all) of the remaining asymptotic ambiguity codified in $\Gamma_k$. Besides, it can be argued that the subsequent description of the entire cosmological system in terms of a complete set of canonical variables would then be optimally adapted to its dynamical behavior from a quantum mechanical point of view. Indeed, with such a criterion, in many physically interesting situations the variables chosen for the gauge-invariant perturbations would undergo no destruction or production of excitations as they evolve.\footnote{This evolution can be formulated with respect to a classical time, as in the context of quantum field theory in a fixed cosmological background, or within hybrid quantum cosmology if one regards one of the homogeneous degrees of freedom as a relational time for the system \cite{hybr-ref}.}. It is worth noting that this lack of self-interactions in the dynamics of the annihilation and creationlike variables can be employed as well to justify the consistency of our choice of normal ordering in the Hamiltonian, given the fact that these variables would then evolve separately, with only a dynamical change in their phase \cite{adiabaticreg1}.

Let us see if the diagonalization of the new MS and tensor Hamiltonians is possible with an adequate choice of $f_k$ and $g_k$. We will do so for the case of the MS variables, but a completely analogous analysis can be performed for the tensor ones by just replacing $s^{(s)}$ by $s^{(t)}$. The desired diagonalization can be attained for all Fourier scales $k\neq 0$ by introducing the function $h_k=g_k^{-1} f_k$ and imposing that, for each $\vec{k}\neq 0$, the interaction terms proportional to ${\bar{a}}_{\vec k}{\bar{a}}_{-\vec k}$ in the Hamiltonian be equal to zero. This, together with the fact that the Hamiltonian vector field $\{.,H_{|0}\}$ is a derivative operator, leads to the following equation:
\begin{align}\label{diagallk0}
k^2+s^{(s)}+h_k^{2}-a\{h_k,H_{|0}\}=0,
\end{align}
where we have used that the canonical condition \eqref{sympl} imposes that $g_k\neq 0$.
What we have obtained is a complex-valued semilinear partial differential equation for $h_k$ that can be recast into a set of first-order ordinary differential equations and has a locally unique solution, as long as the section of initial conditions chosen for them is transversal to the flow of the Hamiltonian vector field $\{.,H_{|0}\}$ \cite{deqs}. Specifically, the characteristic curves of Eq. \eqref{diagallk0} are given by the tuple $[a(r),\pi_{a}(r),\phi(r),\pi_{\phi}(r),h_k(r)]$ where, if the parameter $r$ is identified with the conformal time $\eta$, the background variables in the tuple follow precisely the Hamilton's trajectories dictated by $H_{|0}$ (namely Einstein's equations in the linearized theory), while $h_k(r)$ satisfies:
\begin{align}\label{characteristich}
\frac{dh_k}{dr}=k^2+s^{(s)}+h_k^{2}.
\end{align}

The obtained partial differential equation admits, in principle, many possible distinct solutions that can be associated with different choices of pairs of functions $f_k$ and $g_k$.  In view of this ambiguity, and in order to gain more insight into the dynamical properties of the annihilation and creationlike variables that are allowed by Hamiltonian diagonalization, let us focus on the asymptotic sector of large $k$ and see which restrictions arise for the possible choices of $f_k$ and $g_k$. In doing this, the first step that is needed is the imposition of the asymptotic restriction \eqref{fgasymp}, together with conditions \eqref{xicond} and \eqref{Gammacond}, so that the self-interaction terms become well defined on the vacuum. The resulting coefficients of the factors responsible of the creation of pairs, ${\bar{a}}_{\vec k}{\bar{a}}_{-\vec k}$, have two contributions that, with our assumptions, might give the dominant asymptotic behavior, unless they cancel each other. They are given by the sum
\begin{align}
-\frac{i}{a}f_k\Gamma_k+\frac{i}{4k^3}f_k^2\{s^{(s)},H_{|0}\}.
\end{align}
In particular, the second term arises from the combination of Poisson brackets in Eq. \eqref{Poissoncorr} and the restriction \eqref{xicond} on $\xi_k$. If one wants to eliminate these dominant asymptotic contributions from the self-interaction terms, it is necessary again that the two terms above coincide, except for an opposite sign and up to subdominant corrections. It is possible to see that this pattern repeats order by order in the asymptotic regime $k\rightarrow\infty$, if one keeps on eliminating the dominant contribution at each step and substitutes the result in the considerations at the next order. Actually, it turns out that the dependence of the Hamiltonians on $f_k$ and $g_k$ is such that, in each step, all the terms that might give the dominant contribution are proportional to $f_k^2$ and of known asymptotic order, except a summand proportional to $f_k$ that comes multiplied by the subdominant parts which could not have been fixed before (from the cancellations required in the previous steps), and that therefore remain free.

Taking into account our previous discussion, we make the following {\emph{ansatz}} about the function $\xi_k$ appearing in condition \eqref{fgasymp} for the MS perturbations, concerning its asymptotic expansion:
\begin{align}\label{xiansatz}
\xi_k=-\frac{i}{2k^2}f_k\sum_{n=0}^{\infty}\left(\frac{-i}{2k}\right)^{n}\gamma_n,\qquad \gamma_0=s^{(s)},
\end{align}
where the functions $\gamma_n$ are independent of $k$. Introducing this ansatz in the interaction term of the MS Hamiltonian that is proportional to ${\bar{a}}_{\vec k}{\bar{a}}_{-\vec k}$ [see Eq. \eqref{newMSH}], we arrive at an infinite sum of inverse powers of $k$, up to a global factor $f_k^2k^{-2}$. This asymptotic sum is that of the zero sequence if and only if each of its summands vanishes, something that is achieved by setting
\begin{align}\label{recursion}
\gamma_{n+1}=a\{H_{|0},\gamma_n\}+4s^{(s)}\left[\gamma_{n-1}+\sum_{l=0}^{n-3}\gamma_l \gamma_{n-(l+3)}\right]-\sum_{l=0}^{n-1}\gamma_l \gamma_{n-(l+1)},\qquad \forall n\geq0,
\end{align}
where we have defined $\gamma_{-n}=0$ for all $n>0$. A completely parallel result is obtained for the tensor perturbations, after changing $s^{(s)}$ by $s^{(t)}$. Let us point out that, if the time-dependent masses $s^{(s)}$ and $s^{(t)}$ are $m$-differentiable in all their dependence on the background variables, the iterative equation \eqref{recursion} with initial condition $\gamma_0=s^{(s)}$ is well posed and can be solved, yielding a unique solution $\gamma_n$ for all $n\leq m$.

From our asymptotic analysis above and relation \eqref{fgasymp} it follows that any solution $h_k$ of the semilinear equation \eqref{diagallk0} is such that $(kh_k^{-1}-i)$, if it exists, has an asymptotic expansion in inverse integer powers of $k$ actually equal to $f^{-1}_k\xi_k$, where $\xi_k$ is given by Eqs. \eqref{xiansatz} and \eqref{recursion}. Moreover, if this expansion for $\xi_k$ can be shown to converge (perhaps for certain sets of background trajectories) for sufficiently large $k$, and the resulting function can be uniquely extended --let's say analytically-- to all values of $k\neq0$, then it would uniquely select a solution $h_k$ of Eq. \eqref{diagallk0}.

\subsection{Constraints on the vacuum for the perturbations}

As we have seen, with an adequate choice of annihilation and creationlike variables for the description of the gauge-invariant cosmological perturbations, one can obtain Hamiltonian functions generating their dynamics such that they admit a Fock representation that is diagonal in the basis of $n$-particle states, at least when one considers the asymptotic regime $k\rightarrow\infty$. This is made possible by defining the annihilation and creationlike variables as in Eq. \eqref{constanni} for both tensor and MS perturbations, subject to conditions \eqref{sympl} and \eqref{diagallk0}, with $h_k=g_k^{-1}f_k$. More concretely, the asymptotic behavior of $g_k$ in terms of $f_k$ as a series of inverse powers of $k$ must be given by Eqs. \eqref{fgasymp}, \eqref{xiansatz}, and \eqref{recursion}. These relations, as well as the non-asymptotic Eq. \eqref{diagallk0}, only restrict the functional relation between $f_k$ and $g_k$. However, one can check that, combined with the canonical condition \eqref{sympl}, they asymptotically lead to
\begin{align}\label{fnorm}
|f_k|^2\left[1-\frac{1}{2k^2}\sum_{n=0}^{\infty}\left(\frac{i}{2k}\right)^{2n}\gamma_{2n}\right]=\frac{k}{2},
\end{align}
an identity that properly defines the norm of $f_k$ for sufficiently large $k$. More generally, condition \eqref{sympl} requires that $|g_k|$ be non-zero and that the imaginary part of $h_k$, $\text{Im}(h_k)$, be strictly negative for any values of the canonical variables which describe the homogeneous background. The strict positivity of $|g_k|$ can be imposed without obstructions in our construction, whereas the desired behavior for $\text{Im}(h_k)$ relies on the availabilty of complex solutions to Eq. \eqref{diagallk0} with an imaginary part that is negative at every point. If this is the case, condition \eqref{sympl} in terms of $h_k$ implies that, for all $k\neq 0$,
\begin{align}\label{fnormallk}
2|f_k|^2=-|h_k|^2[\text{Im}(h_k)]^{-1}.
\end{align}
It is worth noticing that, from the fact that the characteristic equation \eqref{characteristich} is of a particularly simple Riccati type, one can easily realize that, at least, solutions $h_k$ with a non-vanishing and negative imaginary part are perfectly attainable by adequately fixing the initial conditions of such equation. Relations \eqref{fnorm} and \eqref{fnormallk} determine the complex norm of $f_k$, both asymptotically and for all $k\neq 0$, once the Hamiltonian diagonalization is achieved. Therefore, after choosing a solution $h_k$ of Eq. \eqref{diagallk0}, the only freedom left in the (asymptotic) definition of the desired annihilation and creationlike variables is the phase of this function $f_k$, phase that we will call $F_k$. We are now in an adequate position to derive explicitly the diagonal expression of the summands in the mode decomposition \eqref{newMSH} of the new MS Hamiltonian. The corresponding formulas for the tensor perturbations are completely similar. A careful computation shows
\begin{align}
{^s{\tilde H}_{|2}}=\frac{1}{a}\sum_{\vec k \neq 0}\Omega_k {\bar{a}}_{\vec{k}} a_{\vec{k}},
\end{align}
where
\begin{align}\label{Omegallk}
\Omega_k=a\{H_{|0},F_k\}-(k^2+s^{(s)})\frac{\text{Im}(h_k)}{|h_k|^2},
\end{align}
which asymptotically behaves as
\begin{align}\label{Omega}
\Omega_k=2\bigg[|f_k|^2+\frac{a}{2}\{H_{|0},F_k\}+\frac{|f_k|^2}{k^2}\sum_{n=0}^{\infty}\left(\frac{i}{2k}\right)^{2n}\big(a\{\gamma_{2n-1},H_{|0}\}+\sum_{l=0}^{n-2}\gamma_{2l+1}\gamma_{2n-(2l+3)}-4s^{(s)}\sum_{l=0}^{n-3}\gamma_{2l+1}\gamma_{2n-(2l+5)}\big)\bigg].
\end{align}

In the spirit of a hybrid quantization of the entire system, for which one should first characterize completely those physical degrees of freedom that are going to be treated with a conventional Fock representation, one can further specify the choice of the phase $F_k$ of $f_k$ with some physical criteria. Firstly, in order to facilitate a quantum representation with nice properties for the functions of the homogeneous variables that are involved in the part of the Hamiltonian constraint that couples to the perturbations, it is rather natural to require that $\Omega_k$ be a positive function of the cosmological background. This is straightforwardly guaranteed if $F_k$ is restricted to be a solution of the following linear partial differential equation
\begin{align}\label{pos}
a\{H_{|0},F_k\}=(k^2+s^{(s)})\frac{\text{Im}(h_k)}{|h_k|^2}+ P_k,
\end{align}
where $P_k$ is an arbitrary positive function of the homogeneous background canonical variables. The remaining freedom in the choice of $F_k$ can then be fixed as follows. When the system is viewed as classical, the MS field ${\mathcal{V}}(\eta,\vec{x})$ (or, similarly, the tensor one) is described by the time-dependent mode coefficients $v_{\vec{k}}(\eta)$ and can be written as
\begin{align}\label{MSfield}
{\mathcal{V}}(\eta,\vec{x})=\sum_{\vec{k}\neq 0}v_{\vec{k}}(\eta)e^{i\vec{k}\vec{x}}=i\sum_{\vec{k}\neq 0}\bar{g}_{k}(\eta)a_{\vec{k}}(\eta)e^{i\vec{k}\vec{x}}+\text{H.c.}
\end{align}
Here, $\vec{x}$ denotes the triple of spatial periodic coordinates adapted to the homogeneity of the background.
Then, in the case under discussion of a(n asymptotic) Hamiltonian diagonalization, the field is given by an infinite sum over all the wave vectors $\vec{k}\neq 0$ of terms that are asymptotically equal to
\begin{align}\label{extr}
\frac{|f_k(\eta)|}{k}\left[1-\frac{1}{2k^2}\sum_{n=0}^{\infty}\left(\frac{i}{2k}\right)^{n}\gamma_{n}(\eta)\right]a_{\vec{k}}(\eta)e^{i\vec{k}\vec{x}-iF_k(\eta)}
\end{align}
and by their Hermitian conjugates. Notice that this decomposition of the field would hold as well if an effective description were available for the entire cosmological system in hybrid quantum cosmology, after replacing all the explicit background dependence of the coefficients of the annihilation and creationlike variables with its effective counterpart and then inserting the genuine effective evolution of the background variables. We thus see that the introduction of a background (or a classical time) dependence in the functions $f_k$ and $g_k$ that define the annihilation and creationlike variables can be understood as the isolation of a part of the dynamical behavior of the field ${\mathcal{V}}(\eta,\vec{x})$ and its identification as a variation that arises from an explicit dependence on the evolving background. The remaining evolution of the field is captured by the time variation of $a_{\vec{k}}(\eta)$ and its Hermitian conjugate. Given the (asymptotic) diagonalization of the Hamiltonian, this latter dynamical transformation becomes just a change of phase,
\begin{align}\label{aevolv}
a_{\vec{k}}(\eta)=e^{-i\int_{\eta_0}^{\eta}\, d\eta' \Omega_k(\eta')}a_{\vec{k}}(\eta_0),
\end{align}
where $\eta_0$ is an arbitrary initial time. We note that this evolution remains valid as well for the corresponding annhiliation operator when it is considered at the level of quantum field theory in our classical cosmological background, since each of the summands in the Hamiltonian are proportional to the number operator and therefore time ordering is trivial. In this context, the freedom in the choice of $F_k$ may even be employed to make the evolution of the annihilation and creationlike variables equal to the identity, something that would render useless the physical information that they carry about the dynamics of the gauge-invariant perturbations. Rather, we can use this freedom to reduce to the minimum the part of the dynamics that we isolate from the MS field ${\mathcal{V}}(\eta,\vec{x})$, thus respecting as much as possible the original dynamical behavior of the gauge invariants, while still allowing for an asymptotic diagonalization of the Hamiltonian. For this, note that the evolution of the complex coefficients $v_{\vec{k}}(\eta)$ is completely characterized by that of their complex norm and phase and, as we have seen, Hamiltonian diagonalization requires a factorization of these coefficients of the form displayed in Eq. \eqref{MSfield}, with $g_k$ of non-trivial norm. It follows that the mentioned criterion for the choice of $F_k$ is satisfactory if such a factorization of the background dependence of the MS field only affects the norm of its complex coefficients $v_{\vec{k}}(\eta)$, but not their dynamical phase. This amounts to require that $g_k$ be real (up to perhaps a constant phase), and hence one must choose (up to a constant)
\begin{align}\label{fphase}
F_k(\eta)=\text{arg}[h_k]+\frac{\pi}{2}=\text{arg}{\left[1-\frac{1}{2k^2}\sum_{n=0}^{\infty}\left(\frac{i}{2k}\right)^{n}\gamma_{n}(\eta)\right]},
\end{align}
where the last equality holds asymptotically and $\text{arg}[.]$ denotes the phase of the complex argument inside the square brackets. With this choice of the phase $F_k$ one can check that
\begin{align}
a\{F_k,H_{|0}\}=\text{Im}(h_k)-(k^2+s^{(s)})\frac{\text{Im}(h_k)}{|h_k|^2},
\end{align}
by using that the complex first order semilinear differential equation \eqref{diagallk0} is equivalent to the following coupled pair of real equations:
\begin{align}\label{imre1}
&k^2+s^{(s)}+\text{Re}(h_k)^2-\text{Im}(h_k)^2-a\{\text{Re}(h_k),H_{|0}\}=0, \\\label{imre2}&
2\text{Re}(h_k)\text{Im}(h_k)-a\{\text{Im}(h_k),H_{|0}\}=0,
\end{align}
where $\text{Re}$ denotes again the real part. Inserting this result in Eq. \eqref{Omegallk} straightforwardly leads to the following function $\Omega_k$ that provides the diagonal Hamiltonian for the perturbations:
\begin{align}\label{OmegaF}
\Omega_k=-\text{Im}(h_k),
\end{align}
which is manifestly positive according to the canonical restriction \eqref{sympl}.

\section{Beyond an asymptotic characterization: Examples}\label{sec4}

As remarked in the previous section, the Hamiltonian diagonalization together with the proposed choice of the phase $F_k$ provides a criterion that fixes completely the asymptotic definition of the annihilation and creationlike variables in the sector of large $k$. It would be desirable to extend these arguments beyond that sector, to cover all the possible wave vectors $\vec{k}$ and then eliminate all the ambiguity affecting the choice of dynamical variables for the gauge-invariant perturbations. Indeed, recall that we have seen that the complete diagonalization of the Hamiltonian for the perturbations is reached with a choice of $f_k$ and $g_k$ subject to conditions \eqref{sympl} and \eqref{diagallk0}. The latter is a partial differential equation with, in principle, an infinite number of different solutions depending on the initial data chosen for $h_k$ at a given set of values of the background variables, data that can even vary from one wave number $k$ to another. Therefore, the choice of annihilation and creationlike variables that diagonalize the Hamiltonian for all $k\neq 0$ does not seem fixed a priori, even if the phase $F_k$ is selected as explained above. Without further knowledge of the analytic expression of the solutions of Eq. \eqref{diagallk0}, it seems natural to restrict our attention to solutions $h_k$ that at least admit an asymptotic expansion of the studied form in inverse powers of $k$, given the asymptotic characterization of the annihilation and creationlike variables that we have obtained under rather mild assumptions on the behavior of $f_k$ and $g_k$. As it was argued in the previous section, if the asymptotic series for $\xi_k$ turns out to converge, the result can be analytically extended to all $k\neq 0$, and the resulting imaginary part of $h_k$ is strictly negative for all $k$, then, by declaring that $\xi_k$ equals $f_k(kh_k^{-1}-i)$ and fixing $F_k$ as in Eq. \eqref{fphase}, we can determine a specific choice of annihilation and creationlike variables for the gauge-invariant perturbations, with an associated Hamiltonian that is diagonal and positive.

It is clear that the selection of a set of annihilation and creationlike variables for the tensor and MS perturbations determines uniquely a Fock representation (and thus a vacuum) for the resulting dynamical variables, in the context of hybrid quantum cosmology. In fact, it also selects a unique vacuum for the original MS field ${\mathcal{V}}(\eta,\vec{x})$ (and for the original perturbative tensor field), when this is viewed as a field defined in a given FLRW background. This situation can be modelled in our system by ignoring all possible backreaction on the homogeneous canonical variables, and adopting for them the dynamics generated by $H_{|0}$. In this context, a decomposition like Eq. $\eqref{MSfield}$ and a choice of initial time, $\eta_0$, completely fix a Fock representation for the field, by simply declaring that $a_{\vec{k}}(\eta_0)=A_{\vec{k}}$ are the coefficients that must be promoted to the annihilation operators that uniquely characterize the vacuum. Moreover, if these annihilationlike variables (and then their conjugate creationlike variables) only undergo a dynamical change of phase, something that amounts to evolve with a diagonal Hamiltonian, the choice of Fock representation becomes essentially independent of the value of $\eta_0$. This is because, in that case, the ratio between the annihilation coefficients $A_{\vec{k}}$ defined with two different values of the initial time is just a constant phase, something that does not affect the choice of vacuum. In the next two subsections, we will show that our asymptotic construction of annihilation and creationlike variables that diagonalize the Hamiltonian actually leads to natural choices, that can be considered standard in quantum field theory, for two examples where the equations of motion of the field are highly symmetric: when the masses $s^{(s)}$ and $s^{(t)}$ are constant, and when the background cosmology is de Sitter spacetime.

\subsection{Constant mass}

Let us study the case in which the masses that appear in the Hamiltonians \eqref{MSham} and \eqref{tensham} of the gauge-invariant cosmological perturbations are positive constants. Notice that, if that is the case, both Hamiltonians reduce to the Hamiltonian of a Klein-Gordon field minimally coupled to a flat spacetime, with a squared mass given by $s^{(s)}$ or by $s^{(t)}$, depending on the type of perturbation that one is considering. The simplest solution to the differential equation \eqref{diagallk0} with a strictly negative imaginary part   is, in this case,
\begin{align}\label{consth}
h_k=-i\sqrt{k^2+s^{(s)}},
\end{align}
which, taking into account restriction \eqref{fnormallk}, leads to
\begin{align}\label{constfg}
g_k=e^{iF_k}\frac{i}{2|f_k|}, \qquad |f_k|^2=\frac{1}{2}\sqrt{k^2+s^{(s)}}.
\end{align}
Since the norms of $f_k$ and $g_k$ turn out to be constant, following the kind of physical motivations that we have explained above, we set $F_k$ also equal to a constant. A completely similar result follows for the tensor perturbations. The corresponding constant coefficients $f_k$ and $g_k$ are well defined $\forall  k\neq 0$ since the mass terms $s^{(s)}$ and $s^{(t)}$ are positive, and they clearly yield the standard Poincar\'e invariant vacuum up to the global phase $F_k$. As we will see in the following, the choice \eqref{consth} of a particular solution of Eq. \eqref{diagallk0} arises naturally from the convergence of the series that determine the asymptotic characterization of the annihilation and creationlike variables that diagonalize the Hamiltonian, derived in the previous section. Indeed, the recursion relation \eqref{recursion} simplifies for constant $s^{(s)}$:
\begin{align}\label{constrec}
\gamma_{n+1}=4s^{(s)}\left[\gamma_{n-1}+\sum_{l=0}^{n-3}\gamma_l \gamma_{n-(l+3)}\right]-\sum_{l=0}^{n-1}\gamma_l \gamma_{n-(l+1)},\qquad \gamma_0=s^{(s)},
\end{align}
which can be easily solved with the introduction of the generating function $J(x)=\sum_{n=0}^{\infty}\gamma_n x^n$. The result is a quadratic equation for $J(x)$ with solution
\begin{align}
J(x)=\frac{1}{2x^2}\left[(1-4s^{(s)}x^2)^{-1/2}-1\right]=\sum_{n=0}^{\infty}\frac{(2n+1)!!\left[4s^{(s)}\right]^{n+1}}{2^{n+2}(n+1)!}x^{2n},
\end{align}
where the latter should be regarded as an asymptotic expansion. From this expression, we conclude that
\begin{align}
\gamma_{2n}=\frac{(2n+1)!!\left[4s^{(s)}\right]^{n+1}}{2^{n+2}(n+1)!},\qquad \gamma_{2n+1}=0,\qquad \forall n\geq 0.
\end{align}
Relations \eqref{fgasymp} and \eqref{xiansatz}, together with restriction \eqref{fnorm} for the norm of $f_k$, lead then to the result \eqref{constfg}, after identifying the asymptotic series appearing in $|f_k|^2$ with the function that has the same power series, namely a square root, and then extending our considerations from the asymptotic region to all $k\neq 0$.  This result was already expected, and it can be straightforwardly deduced from the inspection of the Hamiltonians \eqref{MSham} and \eqref{tensham}, that remain unchanged if $f_k$ and $g_k$ are not background-dependent and that are readily diagonalized for constant mass with a choice like that in Eq. \eqref{constfg}. Nevertheless, this situation serves as an example where we can explicitly check the validity of our asymptotic characterization of a preferred family of annihilation and creationlike variables. Besides, as we have seen, it serves to fix a particular preferred solution of Eq. \eqref{diagallk0}.
 
\subsection{The de Sitter background}

For de Sitter spacetime, the scale factor of the homogeneous and isotropic cosmology with flat spatial slices presents a characteristic evolution $a=-\eta^{-1}H_{\Lambda}^{-1}$, where $\eta$ is the conformal time and $H_{\Lambda}$ is the constant Hubble parameter \cite{mukh}. In our background cosmology, given by an FLRW model minimally coupled to a scalar field with potential $V(\phi)$, one can check that a de Sitter solution can be obtained by setting this potential equal to the constant $3H_{\Lambda}^2/(8\pi G)$. The momentum $\pi_\phi$ of the scalar field is then zero on classical solutions, and thus $\phi$ remains constant throughout the evolution. For these classical solutions of the homogeneous background which, when considering the presence of perturbations, correspond to negligible backreaction, the masses for the MS and tensor gauge invariants reduce to
\begin{align}
s^{(s)}=s^{(t)}=-2\eta^{-2}.
\end{align}
Then, introducing $k\sigma(x)=h(\eta)$ with $x=k\eta$, and taking into account that, in the considered context with no backreaction, the Poisson bracket $a\{h_k,H_{|0}\}$ is just  the conformal time derivative of the function $h_k$ of the homogeneous background, Eq. \eqref{diagallk0} reduces to:
\begin{align}
\dot{\sigma}=1-2x^{-2}+\sigma^2,
\end{align}
where the dot denotes derivative with respect to the rescaled time $x$. Defining $\sigma=-\dot{z}/z$, one finally obtains for $z$ the well-known second-order linear differential equation for the gauge invariants in de Sitter \cite{mukh}:
\begin{align}\label{zeq}
\ddot{z}+(1-2x^{-2})z=0,
\end{align}
which has the two complex independent solutions:
\begin{align}
z_{\pm}=e^{\pm ix}\left(1\pm \frac{i}{x}\right).
\end{align}
The general solution is then given by a linear combination of $z_+$ and $z_-$, which is not fixed a priori. 

In view of the result obtained in the case of constant mass and the asymptotic characterization discussed in the previous section, we will now fix a particular solution $z$ (that determines $h_k$) by computing the asymptotic series that follows from Eqs. \eqref{xiansatz} and \eqref{recursion} in this de Sitter background. First of all, it is not hard to solve the recursion relation \eqref{recursion} up to any order $n$ and deduce that
\begin{align}\label{gammadesitter}
\gamma_{3n}=(-2)^{3n+1}\eta^{-(3n+2)},\qquad
\gamma_{3n+1}=(-1)^{3n+1}2^{3n+2}\eta^{-(3n+3)},\qquad \gamma_{3n+2}=0,
\end{align}
for all $n\geq 0$. The asymptotic series for $g_k$ and $|f_k|$ when one imposes Hamiltonian diagonalization [see Eqs. \eqref{fgasymp}, \eqref{xiansatz}, and \eqref{fnorm}] then become linear combinations of geometric series that, for sufficiently large $k$, converge to
\begin{align}\label{fgdesitter}
kg_k= i f_k\frac{k\eta+k^3\eta^3}{k^3 \eta^3 +i},\qquad \frac{2|f_k|^2}{k}=\frac{1+k^6 \eta^6}{k^4 \eta^4+k^6\eta^6}.
\end{align}
Indeed, one can extend the $k$-domain of these functions to all values of the wave number different from zero without obstructions, and identify the resulting $f_k$ and $g_k$ as the desired functions for the definition of the annihilation and creationlike variables. The corresponding function $h_k$ is
\begin{align}\label{hdesitter}
h_k=\frac{1-ik^3\eta^3}{\eta+\eta^3k^2},
\end{align}
which displays an strictly negative imaginary part for all $k>0$ and $\eta\neq 0$.  With the above expression for $g_k$, one can readily check that the self-interaction terms that destroy and produce pairs of excitations in the Hamiltonians ${^s{\tilde H}_{|2}}$ and ${^T{\tilde H}_{|2}}$ vanish for all $\vec{k}\neq 0$. This confirms that our solution \eqref{gammadesitter} of the recursion relation \eqref{recursion} for $\gamma_n$ in de Sitter spacetime is correct. Actually, it is easy to check that the function $h_k$, given in Eq. \eqref{hdesitter}, coincides with the solution of Eq. \eqref{diagallk0} that arises from choosing $z_+$ as the particular solution of the associated equation \eqref{zeq}. From the local uniqueness of the solutions to that equation, we can thus conclude that our asymptotic characterization of the Hamiltonian diagonalization serves again, in the de Sitter case at hand, to pick out a solution  for all Fourier scales $k$.

After fixing $F_k$ as the argument of $h_k$  in Eq. \eqref{hdesitter} (up to a constant), according to our criteria of positivity of the Hamiltonian and maximal preservation of the original field dynamics in the evolution of the annihilation and creationlike variables, the resulting expansion for the MS field in a de Sitter background turns out to be
\begin{align}\label{MSfieldDeSitter}
{\mathcal{V}}(\eta,\vec{x})=\sum_{\vec{k}\neq 0}\sqrt{\frac{1}{2k}}\sqrt{1+\frac{1}{k^2\eta ^2}}a_{\vec{k}}(\eta)e^{i\vec{k}\vec{x}}+\text{H.c.},
\end{align}
where $a_{\vec{k}}(\eta)$ just oscillates in time as [see Eqs. \eqref{aevolv} and \eqref{OmegaF}]:
\begin{align}
a_{\vec{k}}(\eta)=e^{-ik(\eta-\eta_0)}e^{i[\arctan{(k\eta)}-\arctan{(k\eta_0)}]}a_{\vec{k}}(\eta_0).
\end{align}

One immediately identifies here the standard Fock representation of the MS field on a de Sitter spacetime that defines the Bunch-Davies vacuum (up to, perhaps, a global constant phase in the basis of solutions chosen for the field) \cite{BD}. A completely similar result is obtained for the tensor perturbations.

It is worth noticing that, when the inflaton potential is taken to be the constant $3H_{\Lambda}^2/(8\pi G)$, the de Sitter background solutions that we have investigated in the context of quantum field theory in curved spacetimes actually correspond to certain characteristic curves of Eq. \eqref{diagallk0}, supplemented with the Riccati equation \eqref{characteristich} for $h_k$, if one identifies the characteristic parameter $r$ with the conformal time $\eta$. Therefore, the result obtained in this section should be generalizable beyond the context of classical fixed backgrounds to the entire canonical (truncated) cosmological system treated by the hybrid approach, in the case of a constant inflaton potential.

\section{Adiabatic states}\label{sec5}

As we explained in the previous section, the characterization of a choice of annihilation and creationlike variables with a diagonal Hamiltonian for the gauge-invariant perturbations, introduced as part of the process to construct a hybrid quantization of a primordial cosmology, immediately selects a vacuum for these perturbations when they are viewed as fields propagating on a homogeneous classical cosmology that evolves according to $H_{|0}$ without backreaction. Besides, we have seen that, for some highly symmetric situations and with the choices inspired by our asymptotic characterization, the resulting vacuum coincides with well-known invariant ones that are standard in quantum field theory. Then, for a general evolution of the background FLRW geometry, it would be certainly interesting to compare this Fock representation, that asymptotically diagonalizes the Hamiltonian, with some other choices that can be considered conventional for FLRW universes: the adiabatic states.

These states are typically introduced by looking for approximate solutions $\psi_k$ of the evolution equation of the mode coefficients $v_{\vec{k}}$ of (e.g.) the MS field,
\begin{align}\label{MSeq}
v_{\vec{k}}^{''}+[k^2+s^{(s)}]v_{\vec{k}}=0,
\end{align}
where the prime denotes the derivative with respect to the conformal time, and such that they satisfy the normalization condition
\begin{align}\label{normcond}
{\bar\psi}_k \psi^{'}_k-{\bar\psi}^{'}_k \psi_k=i.
\end{align}
This implies that $\psi_k$ and its complex conjugate would be linearly independent if they were exact solutions of the real equation \eqref{MSeq}. It is easy to convince oneself that this condition restricts the form of $\psi_k$ to
\begin{align}\label{psiw}
\psi_k(\eta)=\frac{1}{\sqrt{2W_k(\eta)}}e^{i\int_{\breve\eta}^{\eta}d{\tilde\eta}\, W_k(\tilde\eta)}, \qquad W_k(\eta)\in \mathbb{R}^{+},
\end{align}
with $\breve\eta$ some arbitrary time that can be set equal to the initial time $\eta_0$ without loss of generality (it only introduces an irrelevant time-independent global phase). This normalization condition is just a specific dynamical realization of relation \eqref{sympl} for the canonical Poisson algebra, inasmuch as it allows one to write \cite{adiabatic}
\begin{align}\label{adiabd}
{\mathcal{V}}(\eta,\vec{x})=\sum_{\vec{k}\neq 0}\frac{1}{\sqrt{2W_k(\eta)}}e^{-i\int_{\eta_0}^{\eta}d{\tilde\eta}\, W_k(\tilde\eta)}a_{\vec{k}}(\eta)e^{i\vec{k}\vec{x}}+\text{H.c.},
\end{align}
where $a_{\vec{k}}(\eta)$ and its Hermitian conjugate behave indeed as canonical annihilation and creationlike variables. In fact, in the case that $\psi_k$ were an exact solution, these variables would be constant in the time evolution. More generally, they evolve with a linear dynamics of the form
\begin{align}\label{adiaba}
a_{\vec{k}}(\eta)=\alpha_k(\eta,\eta_0)A_{\vec{k}}+\beta_k(\eta,\eta_0){\bar A}_{-\vec{k}},\qquad A_{\vec{k}}=a_{\vec k}(\eta_0),\qquad |\alpha_k|^2-|\beta_k|^2=1.
\end{align}

According to these comments, given $\psi_k$, defined at an initial time $\eta_0$, we completely fix a Fock representation of the MS field in a fixed background by declaring that $A_{\vec{k}}$ and its Hermitian conjugate are the variables to be promoted to annihilation and creation operators. Then, since the Cauchy problem is well posed for the MS equation, in order to fix the representation it suffices to give the initial data at $\eta_0$ for the functions $W_k$ that determine $\psi_k$. The aim of introducing adiabatic states is to find functions $\psi_k$ such that the corresponding Bogoliubov coefficients $\alpha_k$ approach the unit as much as possible (at least in an asymptotic sense), so that each of the $\psi_k$'s can be regarded as an approximate solution to Eq. \eqref{MSeq}. 

The practical need to look for this kind of approximation to the exact normalized solutions of the dynamical equation originates from the fact that this equation becomes a complicated non-linear second-order differential equation for $W_k$ that involves explicitly the evolution of the background cosmology. An adiabatic state of order $2N$, with $N\geq 0$,\footnote{Other conventions for the label of the adiabatic order are also adopted in the literature.} is then defined by means of the choice of $\psi_k$ obtained by solving the dynamical equation for $W_k$ iteratively up to order $N$, with a recursive method designed so that the resulting vacuum resembles the Minkowski one, in conformal time, in the limit $k\rightarrow \infty$ and after all of the background time-dependence has been ignored. It is worth noticing that, then, if this method were to converge in the limit $N\rightarrow\infty$, the functions $\psi_k$ would be exact solutions of the field equation \eqref{MSeq} with the desired Minkowskian ultraviolet properties, and one would be able to characterize a unique adiabatic vacuum of infinite order. More specifically, an adiabatic representation of order $2N$ is completely characterized by a decomposition of the form \eqref{adiabd} and \eqref{adiaba}, together with a choice of initial time $\eta_0$, where $\psi_k$ is fixed by Eq. \eqref{psiw} after evaluating $W_k$ as the $N$th-order solution of the iterative equation \cite{adiabaticLR}
\begin{align}\label{recursad}
\left[W_k^{(2N)}\right]^2=k^2+s^{(s)}-\frac{W^{(2N-2)\prime\prime}_k}{2W^{(2N-2)}_k}+\frac{3}{4}\left(\frac{W^{(2N-2)\prime}_k}{W^{(2N-2)}_k}\right)^2,\qquad W_k^{(0)}=k.
\end{align}
Here, we have employed the notation $W_k^{(2N)}$ to denote the iterative solution of $N$th-order for $W_k$. If the resulting function $W_k^{(2N)}$ is indeed real in all the time intervals $[\eta_0,\eta]$ of interest, then for the state of adiabatic order $2N$, with $N\geq 1$, it can be seen that  $\alpha_k=1+\mathcal{O}(k^{-2N-1})$ and $\beta_k=\mathcal{O}(k^{-2N-1})$, where the symbol $\mathcal{O}(.)$ stands for terms of the asymptotic order of its argument or smaller, in the limit $k\rightarrow\infty$ \cite{adiabaticLR}. It follows that $\psi_k$ can actually be regarded as an approximate solution to the MS equation in this regime, with an asymptotic accuracy that improves with the order of adiabaticity. This result is based on the fact that, assuming that the solutions to Eq. \eqref{recursad} are real, one can introduce the {\emph {ansatz}}
\begin{align}
\left[W_k^{(2N)}\right]^2=\left[W_k^{(2N-2)}\right]^2(1+\epsilon_{2N}),\qquad N\geq 1,
\end{align}
and then substitute it in the iterative equation to study the asymptotic behavior of $\epsilon_{2N}$ when $k\rightarrow\infty$. One gets $\epsilon_{2N}=\mathcal{O}(k^{-2N})$, and therefore $W_k^{(2N)}=\mathcal{O}(k)$. This information is enough to deduce the mentioned behavior of the alpha and beta coefficients and determine how good the adiabatic approximation is compared to the actual solutions of the MS equation (at least asymptotically) \cite{adiabatic,adiabaticLR}.

The main objection to the adiabatic approach is that, for general cosmological backgrounds and fixed $k$, there can exist orders $N\geq 1$ such that $W_k^{(2N)}$, defined as in Eq. \eqref{recursad}, would not be real in the time interval $[\eta_0,\eta]$ of interest. In fact, the general situation would be that, within the considered time interval, the function $W_k^{(2N)}$ can be either real or imaginary depending on the subinterval on which it is evaluated, and therefore it would vanish at some instants of time. If this happens to be the case, the associated function $\psi_k$ would not satisfy the normalization condition \eqref{normcond} at those intervals of time for which $W_k^{(2N)}$ becomes purely imaginary, and therefore the decomposition \eqref{adiabd} would not define canonical annihilation and creationlike variables for such times. One can try to solve this problem by replacing the functions $W_k$ appearing in the definition \eqref{psiw} of $\psi_k$ with their norm, and then proceed with the same kind of adiabatic construction that we have discussed above. However, in this modified approach one is forced to take an ansatz for the solutions of Eq. \eqref{recursad} of the form
\begin{align}
\left | W_k^{(2N)}\right |^2=\left |W_k^{(2N-2)}\right |^2(1+\epsilon_{2N}),\qquad N\geq 1.
\end{align}
It is not hard to see that the substitution of this ansatz leads to an asymptotic behavior for $\epsilon_{2N}$ that depends on whether the iteration process \eqref{recursad} provides an even or odd number of independent complex functions $W_k$ in the $N-1$ previous steps, as well as on the intervals of time in which this happens. In particular, one finds that $\epsilon_{2N}(\eta)=\mathcal{O}(1)$ for those periods of the evolution where the iteration supplies an odd number of complex functions $W_k$. Since the expression of the beta coefficients $\beta_k$ for the dynamics of the annihilation and creationlike variables depends on the behavior of $W_k^{(2N)}$ over all the considered interval $[\eta_0,\eta]$, in general they turn out not to decrease as powers of $k^{-1}$ anymore in the discusssed situation, and therefore the iterative method breaks down. Indeed, for fixed $k$ (even if relatively large), this behavior of $\beta_k$ means that the adiabatic procedure leads in general to a function $\psi_k$ that does not provide an approximate solution to the MS equation. So, even if the quantization may be well defined given the initial data at $\eta_0$, the properties of the basis of solutions for the MS field are not under a proper control. This typically implies an anomalous behavior of the two-point function, especially when one considers its low-$k$ contributions, with respect to what one would naturally expect if the adiabatic approximation did not break down.

\subsection{Hamiltonian diagonalization: adiabatic properties}

We finally want to argue that our proposed choice of vacuum for the MS field, defined in a canonical manner by annihilation and creationlike variables that evolve with a diagonal Hamiltonian, seems free of obstructions that affect the adiabatic states and, at the same time, captures the desired properties that lead to their introduction. On the one hand, the decomposition \eqref{MSfield}, with asymptotic coefficients \eqref{extr} of the annihilationlike variables, clearly corresponds to $\beta_k=0$ and $\alpha_k$ equal to the complex exponential of the frequency $\Omega_k$ integrated in conformal time, with this frequency given asymptotically in Eq. \eqref{Omega}. On the other hand, a simple inspection shows that this conformal frequency displays, in the limit $k\rightarrow\infty$, the same asymptotic behavior as the standard frequency for Minkowski spacetime, namely it approaches the wave number $k$, a property which is one of the guidelines for the proposal of the iterative construction of adiabatic states. Furthermore, since condition \eqref{sympl} [and its consequences \eqref{fnorm} and \eqref{fnormallk} for the norm of $f_k$] guarantees that the annihilation and creationlike variables have the correct Poisson algebra, it follows that the resulting basis of solutions that multiply the annihilation coefficients $A_{\vec{k}}$ in the mode decomposition of the gauge-invariant field are actually normalized according to Eq. \eqref{normcond}, and therefore can be factorized in a norm and a phase of the form \eqref{psiw}. In order to reach this conclusion it suffices to notice that, in the context of quantum field theory in our homogeneous cosmological background, we have that the mode coefficients for the momentum of the field are given by $\pi_{v_{\vec{k}}}=\bar{v}_{\vec k}^{\prime}$, and that if $\psi_k$ are exact solutions of the field equation and the field is expressed in terms of them as
\begin{align}\label{solV}
{\mathcal{V}}(\eta,\vec{x})=\sum_{\vec{k}\neq 0}{\bar\psi}_k(\eta) A_{\vec{k}}e^{i\vec{k}\vec{x}}+\text{H.c.},
\end{align}
then the normalization relation \eqref{normcond} is the necessary and sufficient condition for the coefficients $A_{\vec{k}}$ and their Hermitian conjugates to be annihilation and creationlike. One can then conclude that our proposed vacuum for the MS (or tensor) field, based on a Hamiltonian diagonalization, has all the desirable asymptotic properties that one wants of the traditional adiabatic states. 

More generally, beyond the asymptotic regime of large $k$, the method of Hamiltonian diagonalization naturally provides solutions of the real-valued, second-order differential equation that the frequency $W_k$ of any normalized solution $\psi_k$ of the MS equation must satisfy, namely, the non-iterative version of Eq. \eqref{recursad}. In order to see this explicitly, recall that the set of real-valued equations \eqref{imre1} and \eqref{imre2} for the real and imaginary parts of $h_k$ are equivalent to the complex-valued, first-order semilinear equation \eqref{diagallk0} needed for Hamiltonian diagonalization for all $k$. And this equation is well-posed if one chooses adequately the section for its initial conditions. Taking only those solutions $h_k$ with non-vanishing imaginary part [requirement which, we recall, is necessary for consistency with relation \eqref{sympl}], one can solve Eq. \eqref{imre2} for its real part and introduce it in Eq.  \eqref{imre1}. In this way one obtains precisely
\begin{align}\label{imh}
\text{Im}(h_k)^2=k^2+s^{(s)}-\frac{\text{Im}(h_k)^{\prime\prime}}{2\text{Im}(h_k)}+\frac{3}{4}\left[\frac{\text{Im}(h_k)^{\prime}}{\text{Im}(h_k)}\right]^2,
\end{align}
where the prime now stands for the operation of taking the Poisson bracket $a\{.,H_{|0}\}$. Recall that, in the context of quantum field theory in curved spacetimes (namely, without backreaction), this operation reproduces the conformal time derivative. In this restricted context, one therefore recovers the second-order nonlinear differential equation for the frequency $W_k$ that the adiabatic approach tries to solve following the iterative method \eqref{recursad}. It is worth noticing that this differential equation only depends on the absolute value of the imaginary part of $h_k$. Hence, the requirement of its negativity can be satisfied for all intervals of the evolution. In addition, condition \eqref{sympl}, together with our Hamiltonian diagonalization for all $k\neq 0$, allows us to express the field decomposition \eqref{MSfield} as
\begin{align}\label{MSfieldallk}
{\mathcal{V}}(\eta,\vec{x})=\sum_{\vec{k}\neq 0}\frac{1}{\sqrt{-2\text{Im}[h_k(\eta)]}}a_{\vec{k}}(\eta)e^{i\vec{k}\vec{x}}+\text{H.c.},
\end{align}
where we have chosen the phase $F_k$ as in Eq. \eqref{fphase}. Given the form \eqref{OmegaF} of the resulting diagonal Hamiltonian, this decomposition can be clearly expressed as in Eq. \eqref{solV} (at the level of quantum field theory in curved spacetimes), with a basis of solutions $\psi_k$ of the MS equations that can be written in a form similar to Eq. \eqref{psiw}, with a frequency $W_k$ that is equal to minus the imaginary part of our function $h_k$.

To conclude this section, let us emphasize that, starting with a canonical formulation of the entire cosmological system (at the considered order of perturbative truncation) as requested by the hybrid approach, and imposing as a criterion that: a) the constraints of the system be expressed in terms of a complete set of canonical variables, and b) the  contribution of the gauge-invariant perturbations to the Hamiltonian become manifestly diagonal with respect to their annihilation and creationlike variables, we have reached the complex-valued, semilinear differential equation \eqref{diagallk0}. We have seen that the solutions of that equation, in those classical or effective regimes where one can substitute the Poisson bracket $a\{.,H_{|0}\}$ by the conformal time derivative, actually provide solutions of the real-valued differential equation \eqref{imh} that naturally appears in the context of quantum field theory in curved spacetimes, and which is the equation that the adiabatic approach tries to solve iteratively. The simplicity of the complex equation \eqref{diagallk0}, in comparison to the real one \eqref{imh}, supports the convenience of our line of attack to choose a unique Fock representation for the perturbations, apart from being specially suitable for the hybrid quantization of the entire cosmology. Finally, our asymptotic characterization of the solutions to the semilinear complex equation \eqref{diagallk0}, together with the examples studied in the previous section, shed light on the procedure to select a particular preferred solution, provided that the relevant asymptotic series can be arranged to converge. This, together with the discussed choice of the free phase $F_k$ of the function $f_k$, can serve to fix a specific set of annihilation and creationlike variables for the gauge-invariant perturbations within the canonical formalism that describes the entire cosmological system.

\section{Conclusions}\label{concl}

We have investigated the selection of a set of annihilation and creationlike variables with appealing physical properties for the Fock quantization of gauge-invariant perturbations in the framework of hybrid quantum cosmology. With this aim, we have started from a canonical formulation of a homogeneous and isotropic cosmological model with flat compact hypersurfaces and minimally coupled to a scalar inflaton in the presence of small perturbations in the metric and the scalar field, after the truncation of the action of the system at quadratic order in those inhomogeneities (and anisotropies). Specifically, we have studied the contribution of the physical perturbative degrees of freedom to the Hamiltonian, when these are expressed in terms of perturbative gauge-invariant variables. In the flat case at hand, these variables correspond to the MS and tensor fields. In the context of hybrid quantum cosmology, they are to be expressed as a(n inifinite) linear combination of some set of annihilation and creationlike variables, that are to be promoted to operators and thus define the Fock representation of the perturbations. We have checked that, unfortunately, if the coefficients in these linear combinations are constant numbers, the standard MS and tensor Hamiltonians cannot be defined properly on the dense subspace of $n$-particle states by adopting a normal ordering, a fact that suggests that the choice of gauge-invariant variables for the perturbations needs to be improved if one wants to attain a consistent quantization of the entire system. It is worth noticing, nonetheless, that if the constant coefficients that define the annihilation and creationlike variables satisfy condition \eqref{fgasymp}, which is necessary but not sufficient to reach well-defined Hamiltonians, it would suffice to impose that the quantities $k^{-1/2}\xi_k$ form a square summable sequence over all wave vectors $\vec{k}\neq 0$ to guarantee that the classical linearized dynamics of the gauge-invariant variables can be unitarily implemented on the Fock vacuum, even though the infinitesimal generator of this evolution would not be available yet \cite{unique3,unique3b,fmov}. In fact, in the context of quantum field theory in curved spacetimes, it has been proven that this unitary quantum evolution for the perturbations is still valid if the coefficients of the annihilation and creationlike variables in the linear combinations that define the gauge-invariant fields are allowed to depend on time or on background functions, provided that the sequence formed by $k^{-1/2}\xi_k$ is indeed square summable \cite{modeunit,bianchi}. Furthermore, all Fock representations obtained with annihilation and creationlike variables that satisfy this square summability condition turn out to be unitarily equivalent.

In view of the problems found with the Hamiltonians of the perturbations, we have looked for other alternative gauge-invariant annihilation and creationlike variables, with a dynamical behavior different from the original one for the MS and tensor variables since they are defined by means of linear canonical transformations of the latter that depend on the homogeneous cosmological background via two functions $f_k$ and $g_k$. Expressing the total Hamiltonian in terms of this new set of perturbative variables, and of perturbatively corrected homogenous ones that complete a canonical set for the entire system (together with an Abelianized version of the linear perturbative constraints and canonical momenta for them), one easily identifies the infinitesimal generators of the classical dynamics of our new annihilation and creationlike variables. After adopting a mild assumption on the homogeneous background dependence of the canonical transformation that defines these variables, in the asymptotic regime $k\rightarrow\infty$, we have then arrived at the necessary and sufficient conditions that must be fulfilled in order that these new Hamiltonians become properly defined on the Fock vacuum, with normal ordering. These conditions lead to an specific asymptotic behavior (at dominant order) for the background-dependent functions that define the annihilation and creationlike variables, in terms of the time-dependent mass that appears in the MS and tensor Hamiltonians. Such behavior actually guarantees that, according to our comments above, all the acceptable families of annihilation and creationlike variables provide unitarily equivalent Fock representations. Still so, there is an infinite freedom in selecting a representative of this family of representations to describe the primordial vacuum state for the cosmological perturbations. In part with the motivation of removing this ambiguity, we have tried and shown that it is possible to fully determine an asymptotic expansion, when $k\rightarrow\infty$, for the coefficients that define the annihilation and creationlike variables so that the resulting Hamiltonians do not include any self-interaction term in that asymptotic regime, and hence act diagonally on the $n$-particle states. More generally, we have proven that any allowed canonical transformation that defines annihilation and creationlike variables with classical dynamics generated by a diagonal Hamiltonian for all Fourier scales $k$, must be characterized by functions $f_k$ and $g_k$ such that $f_kg_k^{-1}$ is constrained by the semilinear partial differential equation \eqref{diagallk0} with respect to its background dependence.

Once a solution of such semilinear equation is picked out, the only remaining freedom in the choice of annihilation and creationlike variables is the complex phase $F_k$ of $f_k$, which in turn determines the form of the resulting diagonal Hamiltonian. We have discussed how one can fix this phase in order to arrive at a manifestly positive Hamiltonian, something that leads to a partial differential equation for the phase, which nonetheless is linear. Given the ambiguity in selecting a solution to that equation, we have proposed a physically motivated choice of $F_k$ so that, while still respecting the positivity of the Hamiltonian for the perturbations, minimizes the amount of background dependence introduced by the canonical transformation that defines the annihilation and creationlike variables. Let us recall that, provided a specific solution of Eq. \eqref{diagallk0}, the part of the dynamics of the standard MS and tensor fields that does not correspond to the evolution dictated by the diagonal Hamiltonians, and is therefore extracted as background-dependent factors, is completely specified by the set of annihilation and creationlike variables that we choose.

The differential equation that we have found for Hamiltonian diagonalization of the perturbations admits many different solutions. Thus, in order to specify a family of annihilation and creationlike variables with the desired properties, one needs a criterion to choose a particular solution. On the other hand, the asymptotic (ultraviolet) characterization that we have provided for the functions that define those variables is quite specific as an expansion in inverse powers of $k$. To clarify this issue, we have considered two paradigmatic examples: when the effective mass in the dynamics of the perturbations is a constant, and when the potential of the homogeneous inflaton field is constant. These two scenarios correspond, in the context of quantum field theory in curved spacetimes, to fields that propagate in a Minkwoski or in a de Sitter background, respectively. In both cases, we have shown that the asymptotic characterization obtained with our approach fully specifies a solution of Eq.  \eqref{diagallk0}. For these backgrounds, our prescription leads to a choice of annihilation and creationlike variables that determines, respectively, the standard Poincar\'e-invariant vacuum and the Bunch-Davies vacuum. For less restricted types of homogeneous cosmological backgrounds, which are certainly of interest from a quantum cosmology perspective, it may even be impossible to find analytic solutions of Eq. \eqref{diagallk0}, and a similar complication could affect the specification of the coefficients in the corresponding asymptotic expansion. Even so, we have seen that this equation for Hamiltonian diagonalization, which is first-order and semilinear, amounts in fact to the complicated second-order differential equation \eqref{imh} for the imaginary part of the solutions. In the context of quantum field theory in curved spacetimes, this latter equation is equivalent to the field equations if one considers normalized solutions. Typically, one tries to solve it iteratively, with a method that determines the so-called adiabatic vacua and is known to break down for general cosmological backgrounds. Thus, our results may help in the resolution of this complicated equation, providing suitable initial data for the construction of physically appealing solutions, via the derived asymptotic expansions. Actually, we have seen that our asymptotic determination of the annihilation and creationlike variables seems to possess all the properties that the adiabatic approach guarantees for the choice of vacuum when it is applicable.

We should notice that our canonical formulation for the cosmological system, and the availability of background-dependent transformations that lead to Hamiltonian diagonalization, while maintaining a canonical set of variables for the entire cosmology, strongly relies on a truncation of the action at quadratic order in all of the inhomogeneous perturbations. It would be interesting to investigate whether it is possible to extend the ideas that underly our analysis to higher-order perturbative truncations of the action, something that would probably change the notion of perturbative gauge invariants and even of the feasible requirements on their Hamiltonian dynamics. From a more general point of view, it might be insightful to export our ideas to the full action of General Relativity coupled to matter fields, and try to introduce in this full theory canonical transformations that mix the geometrical degrees of freedom with those for the matter content, so that one connects the dynamical notions of both sectors, in such a way that the resulting Hamiltonian adopts a more convenient form for its quantization. Such a formulation, where the dynamical degrees of freedom for the geometry and the matter fields would be well splitted, might even allow a smooth recovery of the regime of quantum field theory in curved spacetimes \cite{ST1,ST2}, if the full quantum representation is under control and a semiclassical limit can be established.

Finally, while finishing the writting of this paper, we became aware of a recent and related work \cite{GK}. In it, the authors make use of the well-known Ermakov-Pinney equation, as well of an extended phase space formulation in the framework of quantum field theory in curved spacetimes, in order to find a particular canonical transformation of the MS gauge invariants such that the resulting Hamiltonian is explicitly time independent. With such transformation, the Hamiltonian then turns out to be diagonal as well. It would be interesting to investigate the relation between the two approaches (ours and that in Ref. \cite{GK}), when one restricts all the considerations presented here exclusively to the treatment of quantum field theory in a curved classical geometry. Actually, one can readily see that it is possible to arrive at an explicitly background-independent (and thus time-independent) diagonal Hamiltonian in our construction, just by adequately choosing the phase $F_k$ [c.f. Eq. \eqref{Omegallk}]. However, there exist two important conceptual differences between the two discussed approaches. First, here we work within the entire canonical (truncated) cosmological model in order to put forward, in the end, a hybrid quantization for the full system, rather than within an extended phase space for the perturbations. Thus, even though both systems are totally constrained by a generator of global time reparametrizations, in Ref. \cite{GK} the relevant constraint is just the sum of the Hamiltonian for the perturbations and the canonical momentum of the time variable in the extended phase space. In contrast, as a result of our canonical approach, in our case this momentum is replaced in the zero-mode of the Hamiltonian constraint with the constraint of the homogeneous cosmology, $H_{|0}$, which has a non-trivial dependence on the homogeneous degrees of freedom (as dictated by General Relativity). Furthermore, in our work the fundamental criterion for the choice of the sets of annihilation and creationlike variables of the perturbations is rooted at the need of disentangling the information about the background cosmology and the perturbative degrees of freedom, in such a way that the resulting canonical variables are optimal for dynamical considerations. One then first imposes that the Hamiltonian for the perturbations be well defined on the vacuum with normal ordering, and then one further restricts the annihilation and creationlike variables so that the resulting Hamiltonian becomes diagonal. We have characterized completely all the freedom available in this procedure. In particular, we have proven that the choice that we have proposed for the phase $F_k$ leads naturally to a frequency for the diagonal Hamiltonian that is positive. Moreover, although we may choose this frequency as a constant, we can perfectly well allow it to depend on the background provided that all of our physical restrictions are respected.

\acknowledgments

This work was supported by Project. No. MINECO FIS2017-86497-C2-2-P from Spain. The authors are grateful to Santiago Prado and Hanno Sahlmann for discussions. B. Elizaga Navascu\'es and T. Thiemann would like to thank K. Giesel and M. Kobler for conversations about their approach to quantum field theory in cosmological spacetimes that makes use of Ermakov's equation.

\end{document}